\documentclass[showpacs,preprintnumbers,amsmath,amssymb,twocolumn,superscriptaddress]{revtex4}
\usepackage{graphicx}% Include figure files
\usepackage{dcolumn}% Align table columns on decimal point
\usepackage{bm}% bold math
\usepackage{ulem}

\begin{document}

\setlength{\abovedisplayskip}{3pt plus1pt minus1pt}
\setlength{\belowdisplayskip}{3pt plus1pt minus1pt}
\setlength{\textfloatsep}{10pt plus1pt minus1pt}
\setlength{\abovecaptionskip}{0pt}

\draft

\title{Isoscaling behavior in $^{48,40}$Ca+$^{9}$Be
collisions at intermediate energy investigated by HIPSE model}
\thanks{Supported by National Natural Science Foundation of
China (10775168, 10775167, 10405032, 10747163, 10605036),
Shanghai Development Foundation
for Science and Technology (06QA14052, 06JC14082 and 05XD14021),
Major State Basic Research Development Program in China
(2007CB815004), and the Knowledge Innovation Project of
Chinese Academy of Sciences (KJCX3.SYW.N2).}

\author{Y. Fu}
\affiliation{Shanghai Institute of Applied Physics, Chinese
Academy of Sciences, Shanghai 201800, China}
\affiliation{Graduate School of Chinese Academy of Sciences,
Beijing 100080, China}
\author{D. Q. Fang}\thanks{Email: dqfang@sinap.ac.cn}
\author{Y. G. Ma}
\author{X. Z. Cai}
\author{W. D. Tian}
\author{H. W. Wang }
\author{W. Guo}
\affiliation{Shanghai Institute of Applied Physics,
Chinese Academy of Sciences, Shanghai 201800, China}

\date{\today}

\begin{abstract}
The fragment production cross sections for 140 MeV/nucleon
$^{48,40}$Ca$+^{9}$Be reactions have been calculated by the Heavy Ion
Phase Space Exploration (HIPSE) model. Isoscaling behavior is
observed. The isoscaling parameters $\alpha$ and $\beta$ for both
heavy and light fragments from HIPSE model calculations are in good
agreement with the experiment data. Studies show that the potential
parameters in the HIPSE model have very small effect on the isoscaling
parameters. The effect of the excitation energy and evaporation on
the isoscaling behavior has also been discussed.
\end{abstract}

%\PACS
%\end{keyword}

%\end{frontmatter}
\maketitle

Motivated by the importance of the symmetry term in the nuclear
equation of state and astrophysical applications, interest in the
isospin effect has considerably increased. In a series of recent
papers, the scaling properties of cross sections for fragment
production with respect to the isotopic composition of the emitting
systems were investigated by M. B. Tsang
et.~\cite{Tsang1,Tsang2,Tsang3,Xu}. Isotopic scaling, also termed as
isoscaling, has been shown to be a phenomenon existing in many
different types of heavy-ion
reactions\cite{Botvina,Friedman,Souliotis}. It is observed by
comparing the same fragments in two similar reactions that differ
mainly in isospin asymmetry. More precisely, the ratio
$R_{21}=Y_{2}(N,Z)/Y_{1}(N,Z)$ is used, where $Y(N,Z)$ is the
isotope yield obtained in the reactions and 2 denotes the more
neutron-rich system. In particular, if two reactions have
approximately the same temperature but different isospin asymmetry,
such ratio exhibit an exponential dependence on the neutron number
$N$ and atomic number $Z$ of the following form:
\begin{eqnarray}
\label{eq1} R_{21}(N,Z)&=&\frac{Y_{2}(N,Z)}{Y_{1}(N,Z)}=C\exp(\alpha
N +\beta Z)
\end{eqnarray}
where $\alpha$ and $\beta$ are the isoscaling parameters and $C$ is
a normalization constant. In the grand canonical limit, $\alpha$ and
$\beta$ will have the form, $\alpha = \Delta \mu _{n}/T$ and $\beta
= \Delta \mu _{z}/T$ where $\Delta \mu _{n}$ and $\Delta \mu _{z}$
are the difference of the neutron and proton chemical potentials for
two reactions. $T$ is the temperature of the system in MeV. As a
matter of fact, these equations were first obtained in the expanding
emitting-source (EES) model\cite{Tsang1}. This behavior is attributed
to the difference of isospin asymmetry between two reaction systems
in similar temperature. The isoscaling has been extensively studied
in different theoretical frameworks, ranging from dynamical models,
such as the antisymmetrical molecular dynamics (AMD) model \cite{Ono}, to
statistical models, such as the statistical multifragmentation (SMM) model
\cite{Botvina}. These investigations of isoscaling are mainly
focused on light fragment. A few studies on the heavy projectile
like residues in projectile fragmentation and fission fragment have
been reported\cite{Fang,Ma,Tian}. In the studies by the statistical
abrasion (STA) model\cite{Fang}, the isoscaling parameters decrease
with the decrease of $Z$ or $N$ and they are close to zero when $Z$
or $N$ become small. The decreasing trend is consistent with
experimental data, but the value of the isoscaling parameter is much
smaller than the data for small $Z$ or $N$ number. This is due to
the simple and geometrical assumptions used in the model. For better
understanding of the isoscaling behavior for both light and heavy
fragments, the Heavy Ion Phase Space Exploration (HIPSE) model\cite{Lacroix}
will be used in this paper. HIPSE parameter's dependence on the
isoscaling parameters and extraction of the symmetry energy
coefficient from the isoscaling parameters will also be
investigated.

\begin{figure*}
\includegraphics[width=11cm]{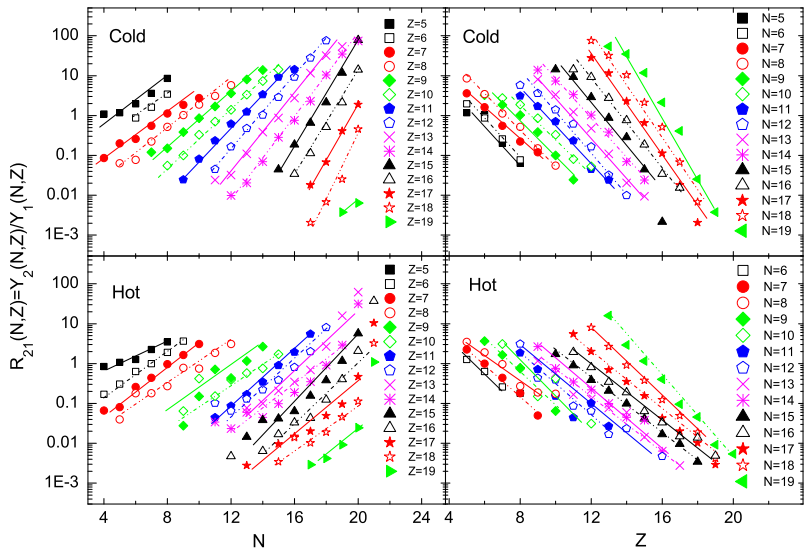}
\caption{Yield ratios $R_{21}(N,Z)$ for hot (lower panels) and cold
(upper panels) fragments from 140 MeV/nucleon $^{48,40}$Ca+$^{9}$Be
reactions versus $N$ for the selected isotopes (left panels)
and $Z$ for the selected isotones (right panels). Different isotopes
and isotones are shown by different symbols. The dashed and solid
lines are drawn to guide the eye.}\label{fig1}
\end{figure*}

Based on the sudden approximation and geometrical hypothesis, HIPSE
model can conveniently simulate heavy-ion interactions at all impact
parameters and thus can constitute a valuable tool for the
understanding of processes such as neck emission or
multifragmentation in peripheral and/or central
collisions\cite{Lacroix}. Accounting for  both dynamical and
statistical aspects of nuclear collisions, HIPSE considers three
different stages of the reaction: the approaching phase, the cluster
formation, and the secondary deexcitation process. For the target and the
projectile nuclei, classical
two-body dynamics of the center of mass is assumed in the entrance channel.
At small distance, a free parameter ($\alpha_{a}$) which defines the hardness
of the potential has been introduced. At the minimal distance of
approach, nucleons in each nucleus are sampled according to a
realistic zero temperature Thomas-Fermi distribution. The
participant and spectator are obtained using simple geometrical
considerations. Nucleons outside the overlap region define the
Quasi-Projectile(QP) and Quasi-Target(QT) spectators. Then two
physical effects, direct nucleon-nucleon collisions and nucleon
exchange, are treated in a simple way. The nucleons in the overlap
encounter a percentage $x_{coll}$ of nucleon-nucleon collisions.
Nucleon exchange is introduced by assuming that a fraction $x_{tr}$ of the nucleons
coming initially from the target (or projectile) and belonging to
the overlap region are transferred to the projectile (or target).
 After these preliminary steps, clusters
are formed using a coalescence algorithm and propagated. To
incorporate the physics of low energy reaction (below the Fermi
energy), two nuclei can fuse. The properties of fusion system are
calculated if two fragments can't separate because their relative
energy is lower than the fusion barrier. This feature leads to
large Final State Interaction (FSI). Once all the FSIs are
processed, the nuclei can't exchange particles anymore and a
chemical freeze-out is reached. After the chemical freeze-out, the
total excitation energy can be determined and shared among
fragments. At this stage, the partition is ready for the
after-burner phase which consists of fragments' propagation  in the
overall coulomb field and secondary decay. The secondary decay is
achieved using the SIMON event generator\cite{Durand}. HIPSE model
only has three adjustable energy dependent parameters($\alpha _{a}$,
$x_{tr}$ and $x_{coll}$). Though HIPSE model is not developed to
describe the projectile fragmentation process, the agreement of the
fragment production cross sections between the calculated results
and experimental data is quite reasonable\cite{Lacroix}.

To study the isoscaling effect in projectile fragmentation, we have
chosen the reaction systems of $^{48,40}$Ca+$^{9}$Be at 140
Mev/nucleon. Values of HIPSE's three energy dependent parameters
$\alpha_{a}=0.55, x_{tr}=0.09, x_{coll}=0.18$ were chosen for beam
energy of 140 Mev/nucleon\cite{Mocko2}. The production cross sections for both
hot (before evaporation) and cold (after evaporation) fragments are
obtained directly from HIPSE model calculations. We extract the
yield ratio $R_{21}(N,Z)$ from the calculated yields $Y_{i}(N,Z) (i=1,2)$ where index 2
refers to the more neutron-rich system. Fig.~\ref{fig1} shows the
yield  ratios of hot projectile-like fragments as a
function of neutron number $N$ for selected isotopes(left lower
panel) and proton number $Z$ for selected isotones (right lower
panel) of the $^{48,40}$Ca+$^{9}$Be reaction systems. The
corresponding ratios for the cold projectile-like fragments are also
shown in Fig.~\ref{fig1} (upper panels). In these figures, different
isotopes and isotones are shown by alternating filled and opened
symbols. The dashed and solid lines are just drawn to guide the eye.

\begin{figure}
\includegraphics[width=6.2cm]{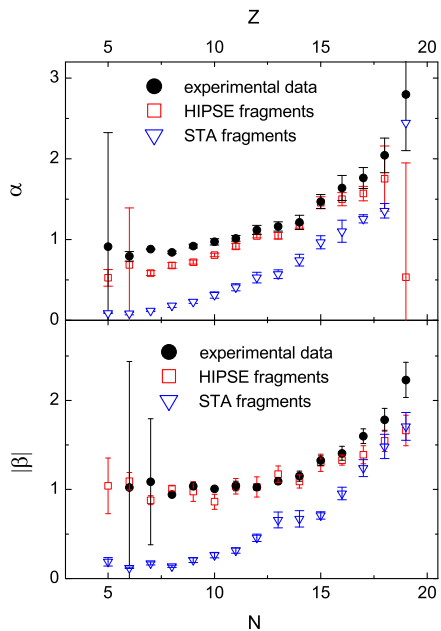}
\caption{Isoscaling parameters $\alpha$ as a function of $Z$ (upper
panel) and $|\beta|$ as a function of $N$ (lower panel) for
140 MeV/nucleon $^{48,40}$Ca+$^{9}$Be reactions. Solid symbols represent the
experimental results with the cross section data taken from
Ref.~\cite{Mocko}. Open squares represent calculations by HIPSE model.
}\label{fig2}
\end{figure}

From the panels of Fig.~\ref{fig1} (hot and cold fragments), we
observe that the ratio for each isotope $Z$ or isotone $N$ exhibits
a remarkable exponential behavior. For each isotope (isotone), an
exponential function of the form $C\exp (\alpha N)(C\exp (\beta Z))$
was used to fit the data. By fitting the calculated points, the
isoscaling parameters $\alpha$ and $\beta$ of hot and cold fragments
are obtained for all isotopes ($Z=5-19$) and isotones ($N=6-19$).
Using the same method, we have also extracted the isoscaling
parameters from experimental fragment production cross section
data\cite{Mocko} and STA model calculations. The isoscaling
parameters from experimental data and cold fragments calculated by
HIPSE and STA model are presented in Fig.~\ref{fig2}. For cold
fragments calculated by STA model, the increase of $\alpha$ and
$\beta$ with increase of $Z$ and $N$ is predicted. Similar trend is
observed in the isoscaling parameters extracted from experimental
data. But the value of the parameter from STA model is much small
than experimental result, especially for light elements. In STA
model, the isoscaling parameters of light fragments is almost
constant for different isotopes because the excitation energy is
almost same for all light fragments in the process of
multifragmentation\cite{Fang}. In Fig.~\ref{fig2}, $\alpha$ and
$|\beta|$ of the cold fragments calculated by HIPSE model show an
increasing trend with the increase of $Z$ or $N$. From the
comparison of the parameters extracted from HIPSE and STA model with
experimental results, it is demonstrated that HIPSE model can
reproduce the isoscaling parameters of the experimental data for
both projectile-like and light fragments quite well, especially for
isoscaling parameter $|\beta|$.

\begin{figure}
\includegraphics[width=7.0cm]{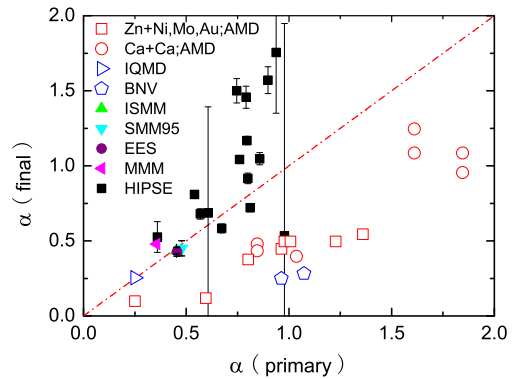}
\caption{Effect of sequential decays on the isoscaling parameter,
$\alpha$(final) and $\alpha$(primary) from different models are shown
in one panel by different symbols. The solid squares are the calculated results
of HIPSE model, other data are taken from Ref.~\cite{Colonna} and references therein.}\label{fig3}
\end{figure}

Isoscaling has been extensively examined in different theoretical
frameworks, ranging from dynamical models to statistical models. In
order to extract symmetry energy information from isoscaling,
detailed understanding of the sequential decay effects on the
isoscaling parameters is necessary. In Fig.~\ref{fig3} the
sequential decay effects of different model are show in one figure.
There are four statistical models, including statistical multifragmentaton
models such as SMM95, ISMM, microcanonical multifragmentation model(MMM)
and the expanding emitting source (EES). The dynamical models used are
the asymmetrized molecular dynamical model (AMD), the Boltzmann-Nordheim-Vlasov
model (BNV) and the isospin quantum molecular dynamical model (IQMD).
As discussed in Ref.\cite{Colonna}, sequential decays effects reduce
the $\alpha$ values in most dynamical models and the reduction are
larger than $50\%$. While the effects are small in statistical
models and its trend is not very clear.
For 140 MeV/nucleon $^{48,40}$Ca+$^{9}$Be reactions calculated by HIPSE model, the
$\alpha$ and $|\beta|$ from hot fragments are much smaller than the
parameters from cold fragments. Sequential decays effects increase
the parameter's values in the model, especially for heavy fragments.
This trend is different from those observed in dynamical and
statistical calculations\cite{Tsang4,Colonna}. The secondary
deexcitation of fragments is described by the SIMON code in HIPSE model\cite{Lacroix}.
In recent paper\cite{Mocko2}, GEMINI\cite{Gemini} replaces SIMON as secondary
decay code in HIPSE model. Results by GEMINI reproduce the cross
sections of the neutron deficient fragments consistently better than
calculations using SIMON. The difference of sequential decays effect
between HIPSE and other models as shown in Fig.~\ref{fig3}
may be mainly due to the evaporation program SIMON used in this model.
A detailed study of the evaporation process between
SIMON and GEMINI, is necessary for a comprehensive understanding of
sequential decays effect on isoscaling parameters in HIPSE model.

\begin{figure}
\includegraphics[width=6.0cm]{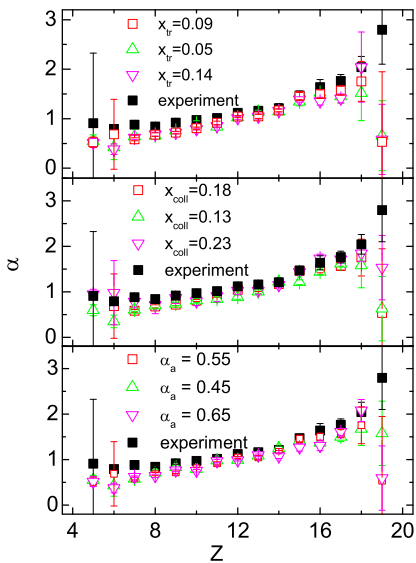}
\caption{HIPSE model parameters' dependence of isoscaling parameter
$\alpha$ for 140 MeV/nucleon $^{48,40}$Ca+$^{9}$Be reactions. From top to
bottom, we present $\alpha$ as a function of $Z$ by independently
adjusting HIPSE parameters $
x_{tr}$(upper panel), $x_{coll}$(middle panel) and $\alpha _{a}$(lower panel).
Experimental data (filled square) with the cross sections taken from
Ref.~\cite{Mocko} are shown in the figures too.}\label{fig4}
\end{figure}

As a phenomenological model, HIPSE contains many input parameters.
But for most of them, a fixed constant value is used in the model.
The HIPSE model only has three adjustable parameters. For the
reaction system $^{48,40}$Ca+$^{9}$Be in this paper, these
parameters have been adjusted by comparing with experimental data
and extrapolated with a simple function to the beam energy of 140
Mev/nucleon\cite{Mocko2}. It is necessary to study the HIPSE's
parameter dependence of isoscaling parameters. In Fig.\ref{fig4}
(from upper to bottom panel), we present the results by adjusting
one of the HIPSE parameters $x_{tr}$ (upper panel), $x_{coll}$
(middle panel) and $\alpha _{a}$ (lower panel) while the other two
HIPSE parameter's values are fixed. The isoscaling parameters
$\alpha$ show the same trend, and their values don't change with the
adjustment, for most fragments. In Fig.~\ref{fig4}, all HIPSE
parameters just exhibit small influence on the isoscaling parameter
for heavy fragment. So HIPSE's parameters dependence of isoscaling
parameter can be negligible in our studies.

\begin{figure}
\includegraphics[width=6.cm]{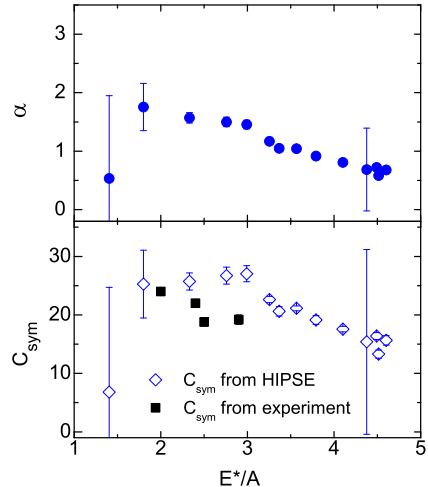}
\caption{The isoscaling parameter $\alpha$ (upper panel) and the
extracted symmetry energy coefficient from $\alpha$ ($C_{sym}$,
lower panel) as a function of excitation energy ($E^{*}/A$) for
140 MeV/nucleon $^{48,40}$Ca+$^{9}$Be reactions. The solid squares are the
experimental data taken from Ref.~\cite{Souliotis}.}\label{fig5}
\end{figure}

In the grand-canonical approximation, assuming same temperature
$T$, the isoscaling parameters $\alpha$ and
$\beta$ are proportional to the difference of the neutron and proton
chemical potentials for two systems. The particular interest comes
from their connection with the symmetry energy coefficient. From the
difference of chemical potentials within the grand-canonical
approximation, it is possible to connect the isoscaling parameter
$\alpha$ to the difference of asymmetry $(Z/A)$ between the two
systems considered, the values of symmetry energy coefficient and
temperature\cite{Tsang4}. The relation is
\begin{eqnarray}
\label{eq2}\alpha =
\frac{4C_{sym}}{T}[(\frac{Z}{A})^{2}_{1}-(\frac{Z}{A})^{2}_{2}]=\frac{4C_{sym}}{T}\Delta
[(\frac{Z}{A})^{2}]
\end{eqnarray}
where $\alpha$ is the isoscaling parameter extracted from the
calculated yields of fragments, $T$ is the temperature,
$(Z/A)_{i}$ is the ratio between nuclear charge and mass number.
The symmetry energy coefficient $C_{sym}$ is directly related to
the symmetry energy (per nucleon) of a given fragment having
asymmetry $(I=(N-Z)/(A))$ $E_{sym}=C_{sym}I^{2}$. In HIPSE model,
the total excitation energy is determined by the total energy
balance and shared among fragments. In upper panel of
Fig.~\ref{fig5}, we present the extracted parameter $\alpha$ as a
function of excitation energy ($E^{*}/A$). $\alpha$ show a
decreasing trend with the increasing of $E^{*}/A$. According to
Eq.~(\ref{eq2}), if we use the Fermi-gas relationship between
excitation energy per nucleon and temperature
$E^{*}/A=\frac{1}{a}T^{2}$ to calculate $T$, with the inverse
level density parameter $a=8-13$ ($a=10$ is used in this paper),
the symmetry energy coefficient ($C_{sym}$) could be extracted.
Results extracted from $\alpha$ are shown in the lower panel of
Fig.~\ref{fig5}. The symmetry energy coefficient from $\alpha$ is
around 20 Mev and decreases with the  increasing of excitation
energy $E^{*}/A$. Similar dependence was also observed in the
experimental data\cite{Souliotis} but their $C_{sym}$ values is a
little bit smaller than HIPSE's result. The experimental studies
of heavy-residue isoscaling reveals the gradual decrease of the
symmetry energy coefficient with increasing excitation energy in
the range$E^{*}/A=2.0-2.9$ \cite{Souliotis,Fevre}, a similar trend
is obtained by HIPSE in a big range $E^{*}/A=2.0-4.5$.

In conclusion, the fragment production cross sections of 140 MeV/nucleon
$^{48,40}$Ca+$^{9}$Be reaction systems have been
studied by HIPSE model. Isoscaling behavior is observed in the ratio
of the fragment yields from the two systems. The extracted
isoscaling parameters from HIPSE model calculations are in good
agreement with recent experimental data for both projectile-like and
light fragments. The excitation energy dependence of symmetry energy
coefficients obtained from the calculations is also consistent with
the experimental data. Sequential decays and parameters of HIPSE
model have some effects on isoscaling parameters, especially for
heavy fragments. Thus HIPSE model could be used to investigate the
isoscaling phenomena of fragments produced in central and peripheral
collisions.

%\footnotesize
{}

\end{document}